# Enhanced Out-of-Band Sensing Algorithm for Cognitive Radio Networks

Yalew Zelalem Jembre and Young-June Choi

*Department of Computer Engineering, Ajou University*
*Woncheon-Dong, Yeongtong-gu, Suwon, Korea*
{zizutg,choiyj}@ajou.ac.kr

*Abstract* – In cognitive radio networks (CRN), Out-of-Band (OoB) spectrum sensing provides seamless communication. Cognitive radio (CR) users, so called secondary users (SUs), should avoid interference with primary users (PUs), the owner of the licensed band, while trying to access the unused licensed or unlicensed band, for spectrum utilization. When PUs request to access their band, SUs need to vacate the band, thus it is inconvenient to provide seamless communication without OoB sensing. In this paper, we suggest an OoB sensing algorithm to guarantee seamless communication and also minimize the interference of SUs on PUs. Also we obtain analysis-based achievable throughput by considering the OoB sensing duration. To verify our algorithm, we perform simulation and find that the effect due to OoB sensing on the aggregate throughput is insignificant.

*Keywords* – Cognitive radio, spectrum sensing, out-of-band (OoB) sensing, throughput analysis

## I. INTRODUCTION

In recent years, owing to the prodigious interest in the use of wireless communication, radio spectrum is getting scarce. Up until this moment, spectrum is assigned statically to a specific licensed service and its users, which caused spectrum sacristy as spectrum is a limited resource. Moreover, this allocation system degrades spectral efficiency [1], and also does not prevent licensed users, i.e., primary users (PUs) from being interrupted by other users [2].

Conversely, a recent study [3] reveals spectrum scarcity occurs due to the inefficient spectrum allocation rather than the actual physical shortage and another study by Federal communication commission (FCC) of US [4] has shown 70% of the allocated spectrum is not utilized. To mitigate this problem, FCC proposed a new spectrum assignment policy called, Dynamic Spectrum Access (DSA) and Cognitive Radio Networks (CRN) for its spectrum allocation mechanism. The IEEE has formed IEEE 802.22 group to define a standard interface PHY and MAC for cognitive radio [5].

Despite many constraints, *spectrum sensing* is the key to successful deployment of a CRN. When SUs want to use any unutilized licensed spectrum/white space, SUs should first detect if PUs are present or not, in order to avoid interference, which is done via spectrum sensing. There are several methods of detecting PUs such as *Energy detection, Cyclostationary feature detection,* and *Matched filter* [6].

SUs should keep monitoring the spectrum currently being in use periodically for the return of the primary users. [5] IEEE 802.22 defined two quite periods to perform this, *fast sensing* and *fine sensing*. In the fast sensing, it takes around several milliseconds and executed periodically, so SUs use it to collect information about the channel that is currently being in use. In the fine sensing, it is only triggered if information collected during fast sensing indicates there is a probability of existence of a PU in the channel. Both periods are used to detect a PU in current channel and they are classified as *In-Band* (IB) sensing.

If PUs' appear in that channel, SUs need to vacate the channel to avoid interference with PUs but SUs' transmission is inevitably interrupted. In such a scenario, SUs either search another available channel right after leaving the current channel, which is called *reactive sensing* but it might take longer time due to fading and shadowing; or explore other channels for availability while using the current one, which is called *proactive sensing*. These are classified as *Out-of-Band* (OoB) sensing which is performed over channels that are not being used currently.

In this paper, we are interested in the proactive OoB sensing. When dual-antenna equipment is available, one antenna is used for data transmission and the other for searching white spaces, but it is expensive and inefficient and even might cause severe interference among antennas, thus degrade SUs' performance [7]. Therefore, we assume that SUs are equipped with a single antenna, where SUs need to sacrifice transmission time to be able to search backup channels. The OoB sensing is performed periodically with the duration equal to fine sensing and the duration can be used for fine sensing or data transmission when there is no OoB sensing.

In literature, there are some approaches to OoB sensing. IEEE 802.22 has suggested fine and fast sensing to detect primary users but there is no indication of proactive or OoB sensing to facilitate the condition [5]. In addition to that, it does not provide any algorithm when to perform the sensing either. In [11], the authors suggest efficient out-of-band cooperative sensing method in the sensing process confirming the appearance of the incumbent user in the channel being used by the CR user. Also in [12], the authors propose a framework for proactive spectrum access and provide detailed prediction methods under the assumption of exponential and periodic traffic models and also propose different prediction and schedule schemes using different sensing capabilities. Both in [11] and [12], even though an algorithm is provided when to perform OoB sensing, it does not include fine and fast sensing and the impact of OoB sensing on the current channel is not presented. In [8], we propose OoB sensing and the probability of discovering an OoB channel and compared the throughput in terms of the number of frames.

In this paper we propose an OoB sensing algorithm that includes how and when to perform fast, fine and OoB sensing. In addition to that, we analyse the throughput of SUs mathematically to compare the impact of OoB or proactive sensing on the IB channel. From simulation results, we prove that the effect of OoB sensing on the aggregate throughput is insignificant.

The remainder of this paper is organized as follows. Section II explains background of spectrum sensing, and Section III presents our proposed sensing scheme. In Section IV, we provide our mathematical analysis on the throughput. In Section V, we show numerical results, and Section VI concludes this paper.

## II. BACKGROUND

### A. In-Band and Out-Of-Band sensing

We consider two types of frames to be used, one that only incorporates IB sensing $\tau$, whereas the other incorporates both IB sensing duration $\tau$ and duration that could either be used for OoB or fine sensing $\omega$ [8]; however, the frame duration is the same T. Fig. 1 shows the frame structures. In addition to that, IB sensing is done on a frame-by-frame basis but OoB and fine sensing are performed when necessary. If there is no need to sense during this duration it can be used for data transmission.

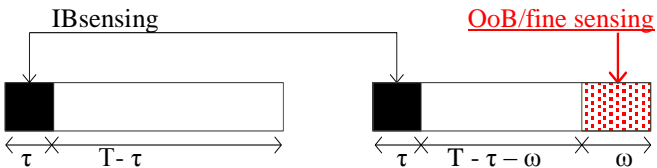

Fig. 1 Frame structure for spectrum sensing

We consider a CRN that consists of a BS and several SUs and cooperative sensing is employed in the network and a BS makes a decision based on data collected from SUs on the channel occupancy [8].

### B. Channel on time

Currently, two frequency bands are available for CRNs, 400-800 MHz (UHF TV bands) and 3-10 GHz. On TV bands, the channel occupancy is nearly fixed and it makes easier to capture and estimate the channel occupancy, 'on' time $T_{on}$ of these bands [9]. But on digital signals with 'on' time $T_{on}$ occupying a certain frequency, estimating the channel occupancy is a bit difficult due to the dynamic characteristic of the channel; although such signals can be captured by fine tuning, sensing parameters and the 'on' time $T_{on}$ of such signals can be estimated from samples of channel occupancy as follows [9]:

$$T_{on} = N_{on} + 2T_s \qquad 1$$

Here, $N_{on}$ is the number of samples taken during 'on' time of signal and can be given as $N_{on} = \tau f_s$ or $N_{on} = \tau f_s$ for IB and OoB sensing, respectively, where $f_s$ is signal sampled at sampling frequency. A factor of $2T_s$ is added to estimate the account of uncertainty and $T_s$ is the sampling time.

### C. Energy detection

Energy detection is one of the methods to detect the presence of PUs in a certain channel. The presence and absence of PUs can be measured at SUs and can be represented by two hypotheses $H_1$ and $H_0$ respectively:

$$y(n) = s(n) + u(n) \ldots \ldots \ldots \ldots \ldots \ldots H_1$$
$$y(n) = u(n), \ldots \ldots \ldots \ldots \ldots \ldots \ldots \ldots H_0$$

where $y(n)$ is the received signal at SUs, $s(n)$ is the transmitted signal by PUs and $u(n)$ is the Gaussian noise. Using the analysis in [10], the test statistics is given as follows:

$$T(y) = \frac{1}{N}\sum_{n=1}^{N}|y(n)|$$

Under a complex valued PSK signal and CSCG noise case, based on the test statistics the probability of detection $P_d$ and the probability of false alarm $P_f$ for a certain threshold $\epsilon$ are respectively given by [10]:

$$P_d(\epsilon,\tau) = Q\left(\left(\frac{\epsilon}{\sigma_u^2} - \gamma - 1\right)\sqrt{\frac{\tau f_s}{2\gamma + 1}}\right)$$

$$P_f(\epsilon,\tau) = Q\left(\left(\frac{\epsilon}{\sigma_u^2} - 1\right)\sqrt{\tau f_s}\right)$$

where $\gamma$ is the SNR of channel of interest and Q(.) the complementary distribution function of standard Gaussian, i.e. [10],

$$Q(x) = \frac{1}{\sqrt{2\pi}}\int_x^{\infty} exp\left(-\frac{t^2}{2}\right)dt.$$

## III. PROPOSED SENSING SCHEME

In the previous section, we have described the sensing duration for different purpose; although IB sensing is done on each frame, OoB sensing and fine sensing are triggered only when needed. Otherwise the duration allocated for OoB or fine sensing could be used for data transmission.

In [5], it has been suggested that fine sensing can be done based on the data collected during IB sensing. However, since OoB sensing is not proposed in [5], we provide an interval of the OoB sensing. And also we resolve the problem that OoB and fine sensing may be triggered at the same time. We propose an algorithm that could handle this situation.

## A. Sensing interval

The interval designed here is only for OoB sensing, the interval of fine sensing is out of the scope of this paper. We assumed that there could at least be one OoB channel from the scanned channel '$i$' ($S_C^i$), $i = 1, 2, 3 \ldots n$ and each channel has a capacity of $C_i$. Applying equation (1), the 'on' time of any channel $S_C^i$, is given as follows; $T_{on}^i = N_{on}^i + 2T_s^i$ and also the 'on' time of current channel is given by $T_{on}^{cur} = N_{on}^{cur} + 2T_s^{cur}$.

The total amount of data that could be transferred on a link could be found by multiplying the 'on' time of a channel by channel capacity. Therefore, data that current channel support $D_{sup}^{cur}$, $D_{sup}^{cur} = T_{on}^{cur} * C_{cur}$ and the data an OoB channel 'i' support $D_{sup}^i$, can be given by $D_{sup}^i = T_{on}^i * C_i$. And, if we denote the total amount of data a SU wish to transfer by, $D_{SU}^{tot}$ then the sensing interval if one of channels $C_i$ become a possible candidate as a backup channel, can be denoted by $S_I^i$. Then the objective of finding an OoB backup channel which gives a relaxed sensing interval becomes identifying a channel which has longer 'on' time than the current channel and that could support maximum of data transfer rate than other channels. Mathematically, the optimization problem can be stated as follows:

$$\max_{S_I^i} S_I^i = \frac{D_{sup}^{cur} + D_{sup}^i}{D_{SU}^{tot}} \qquad 2$$
$$s.t. \qquad T_{on}^i > T_{on}^{cur}$$

The above equation will enable SUs to determine to choose which OoB backup channel $C_i$ could make larger interval between consecutive OoB sensing frames so that the overhead on the aggregated throughput due to OoB sensing could be insignificant.

**Algorithm 1** Obtaining maximum interval $Max_I$
1: Perform OoB sensing on the first frame,
2: Set $Max_I$ =1, $Res_C$ and compute $T_{on}^{cur}$
3: Scan channels $C_i$ , I = 1,2,3…n,
4: **for each** I = 1 **to** n do
5: **if** $C_i$ = 'available' **then**
6: Compute $T_{on}^i$
7: **if** $T_{on}^i > T_{on}^{cur}$ **then**
8: Compute $S_I^i$
9: **if** $Max_{II} < S_I^i$ **then**
10: $Max_I = S_I^i$
11: $Res_C = C_i$
12: increment i and go to line 5
13: **end if**
14: **end if**
15: **end if**
16: **end for**
17: End

## B. Sensing algorithm

Here, we will discuss two important algorithms the first algorithm enables SUs to get a reserve channel $Res_C$, that could give the maximum interval, $Max_I$, which is nothing but max of $S_I^i$, between consecutive OoB MAC frames. To do so, SUs at the start of any communication or during channel switching, can scan the OoB channels, select only the available ones, compute the 'on' time of current and scanned channels, then compare them; then, take the channel as a reserve channel that could give maximum interval between two consecutive OoB frames. The second algorithm will take $Max_I$ as input and regulate when to perform IB, OoB or fine sensing.

**Algorithm 2** Deciding what type of frame to use
1: Take $Max_I$ and $Res_C$ as an input
2: **for each** count = 1 **to** $Max_I$ do
3: **if** count < $Max_I$ **then**
4: **if** fine sensing = 'required' **then**
5: Perform fine sensing
6: **if** FS result = 'vacate' **then**
7: Switch to $Res_C$
8: Call Algorithm 1
9: **else**
10: Stay on $C_{cur}$ Perform IB sensing only
11: Increment count go to line 4
12: **end if**
13: **else** // if fine sensing not required
14: Stay on $C_{cur}$ Perform IB sensing only
15: Increment count go to line 4
16: **end if**
17: **else** //count = $Max_I$ then OoB and FS sensing
18: **if** fine sensing = 'required' **then**
19: Schedule OoB sensing for $Max_I$ + 1[th] frame
20: Perform fine sensing
21: **if** FS result = 'vacate' **then**
22: Switch to $Res_C$
23: Call Algorithm 1
24: **else if** FS result ≠ 'vacate' **then**
25: Stay on $C_{cur}$ Perform OoB sensing on $Max_I$+1[th]
26: Call Algorithm 1
27: **end if**
28: **else** // if fine sensing not required
29: Stay on $C_{cur}$ Perform OoB sensing only
30: Call Algorithm 1
31: **end if**
32: **end if**
33: **end for**
34: End

Therefore, since the algorithm keeps two consecutive OoB frames as far apart as possible, there will be less effect of OoB frames on the aggregate throughput and since SUs maintain a reserved channel it would help to avoid interference with PUs. Most importantly, SUs will keep seamless communication because of the reserved channel.

## IV. THROUGHPUT ANALYSIS

In the previous sections, we have described how sensing should be done in order to avoid the interference and provide seamless communication. In this section, we will obtain the achievable throughput and the aggregated throughput for the IB channel, based on the analysis and scenarios given in [10]. We consider the concept of including OoB sensing and approximation of the aggregated throughput.

There are two possible throughputs, when SUs operate on PUs' spectrum. When SUs operate in the presence and absence of PUs, the throughput is denoted by $C_1$ and $C_0$, respectively. There are four scenarios SUs can operate at PUs band and the achievable throughput under each scenario is as follows. In scenario 1, there is no PU, false alarm does not exist and there is no OoB sensing, so throughput is $\frac{T-\tau}{T}C_0$. In scenario 2, there is no PU and no false alarm is generated but OoB sensing is required, so throughput is $\frac{T-\tau-\omega}{T}C_0$. In scenario 3, there is a PU but it is not detected and no OoB sensing is held, so throughput is $\frac{T-\tau}{T}C_1$. In scenario 4, there is a PU but it is not detected and OoB sensing is held, so throughput is $\frac{T-\tau-\omega}{T}C_1$.

For a certain band, suppose the probability that a PU is active is $P(H_1)$ and the probability PU is inactive is $P(H_0)$, which implies $P(H_0) + P(H_1) = 1$. Then the probability for the first and second scenarios to happen is given by $(1 - P_f(\epsilon,\tau))P(H_0)$; and the probability for third and fourth scenario to happen is given by $(1 - P_d(\epsilon,\tau))P(H_1)$. Using [10], If we define, for each scenario,

$$R_{0,1}(\epsilon,\tau) = \frac{T-\tau}{T}C_0\left(1 - P_f(\epsilon,\tau)\right)P(H_0), \quad 3$$

$$R_{0,2}(\epsilon,\tau + \omega) = \frac{T-\tau-\omega}{T}C_0\left(1 - P_f(\epsilon,\tau)\right)P(H_0), \quad 4$$

$$R_{1,3}(\epsilon,\tau) = \frac{T-\tau}{T}C_1\left(1 - P_d(\epsilon,\tau)\right)P(H_1), \quad 5$$

$$R_{1,4}(\epsilon,\tau + \omega) = \frac{T-\tau-\omega}{T}C_1\left(1 - P_d(\epsilon,\tau)\right)P(H_1). \quad 6$$

When using IB frame $R_{IB}$ and OoB frame $R_{OoB}$ and applying equations (3), (4), (5) and (6), the average throughput of SUs is given by

$$R_{IB}(\tau) = R_{0,1}(\epsilon,\tau) + R_{1,3}(\epsilon,\tau), \quad 7$$

and

$$R_{OoB}(\tau + \omega) = R_{0,2}(\epsilon,\tau + \omega) + R_{1,4}(\epsilon,\tau + \omega), \quad 8$$

respectively. There is no need to include the threshold for OoB channels for calculating the throughput since it does not affect the quality of IB channels.

Even though it is possible to achieve some throughput while PU exists, as in third and fourth scenario, it is quite insignificant. Therefore to get aggregate throughput we will only use the first two scenarios. Letting the numbers of total frames be $F_{no}^{tot}$, it is obtained by dividing 'on' time of the current channel by frame duration $F_{no}^{tot} = \frac{T_{on}^{cur}}{T}$. Then if OoB frame is used, the aggregate throughput of the network, $R_{agg}$ can be obtained from (3) and (4), and also is given by:-

$$R_{agg} = \sum_{i=0}^{F_{no}^{IB}} R_{0,1}(\epsilon,\tau) + \sum_{i=0}^{F_{no}^{OoB}} R_{0,2}(\epsilon,\tau + \omega). \quad 9$$

However, if there is no OoB sensing, all frames will be used by IB frames only and the aggregate throughput can be given as follows, only using(3):

$$R_{agg} = \sum_{i=0}^{F_{no}^{tot}} R_{0,1}(\epsilon,\tau), \quad 10$$

where $F_{no}^{OoB}$ is the number of OoB frames and $F_{no}^{IB} = F_{no}^{tot} - F_{no}^{IB}$ is the number of IB frames used during transmission.

V. NUMERICAL RESULTS

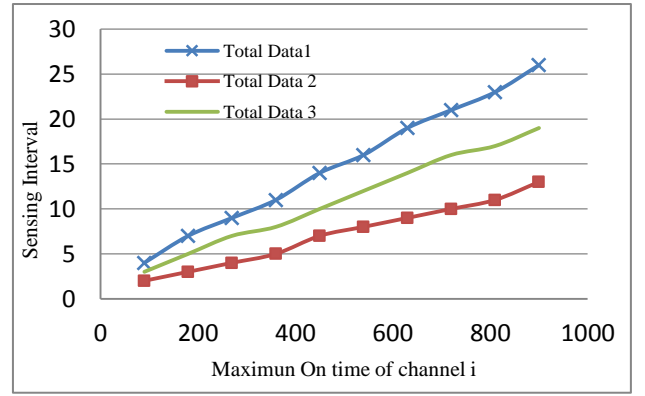

Fig. 2 Sensing interval in terms of maximum on-time

In order to validate the performance of the proposed scheme we present numerical results. We have taken the following assumptions to perform the analysis. Correspondingly, the probability of detection and false alarm is given by 0.9 and 0.1. The Channel capacity for $C_0$ and $C_1$ is taken to be 6.6582 and 6.6137 respectively. The probability of $H_0$ is 0.9 and the probability of $H_1$ is 0.1. Frame duration is taken to be 100msec.

In Fig. 2, we compare the sensing interval for different sets of data. It is shown that as the maximum on-time that could be found from OoB channels increases, the sensing interval between consecutive OoB sensing will also increase so the effect of OoB sensing specially on large data transfer is low.

In Fig. 3, the average achievable throughput is compared, when the OoB sensing is used and unused. The OoB sensing interval is varied while keeping the IB sensing interval to 1msec. It is obvious the throughput when OoB sensing is used shows a small amount of degradation but numerically the difference is less than 0.1 compared to the throughput achieved without OoB sensing.

Since the achievable throughput that could be gained when a PU is on but not detected is very less; in Fig. 4 we compare the exact achievable throughput when there is no PU and it shows almost similar result as compared to results in Fig. 3 which confides our assumption perfectly.

Finally, Fig. 5 shows the effect of OoB sensing on the aggregate throughput. Since the achievable throughputs show 1msec for IB sensing and it is recommended by 802.22 working group, we only use 1msec for IB sensing. As seen in Fig. 5, the total on-time of the current channel is varied and almost all the time the aggregate throughput with and without employing OoB sensing is the same.

## VI. CONCLUSION

In this paper we have proposed OoB sensing to improve the performance of cognitive radio to look for OoB channels for later use while using the current channel. OoB sensing is a means to improve seamless communication while avoiding interference. Our analysis confirms that the effect of OoB sensing on throughput is insignificant.

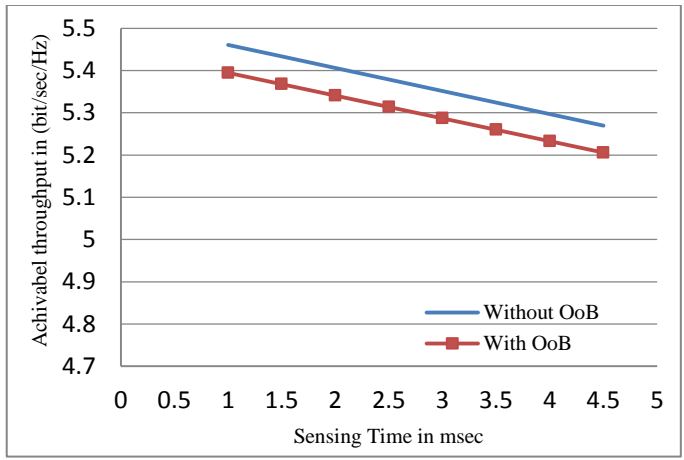

Fig. 3 The average throughput with and without OoB sensing when a PU appears

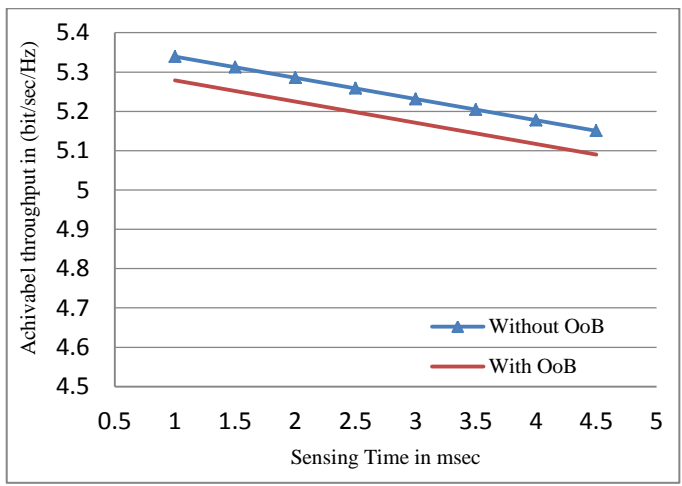

Fig. 4 The average throughput with and without OoB sensing when no PU appears

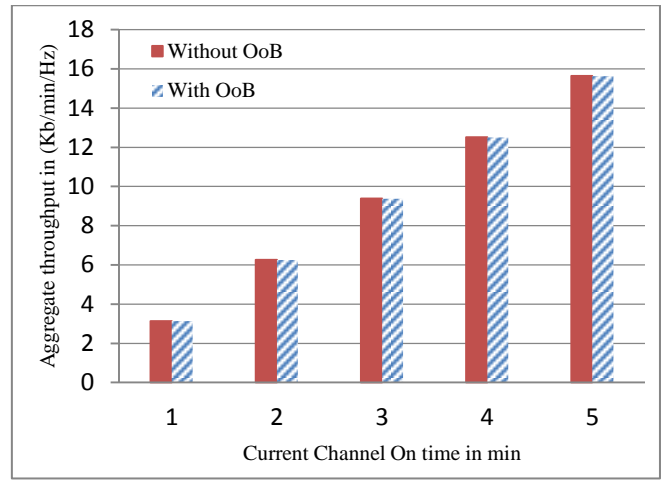

Fig. 5 Average throughput in terms of channel on-time

ACKNOWLEDGMENT

This research was supported by the MKE (The Ministry of Knowledge Economy), Korea, under the ITRC(Information Technology Research Center) support program supervised by the NIPA(National IT Industry Promotion Agency" (NIPA-2011-(C1090-1121-0011)).